\documentclass{aastex63}
\usepackage{amsmath}    
\numberwithin{equation}{section}
\usepackage{xcolor}

\newcommand{\vect}[1]{\boldsymbol{#1}}

\newcommand{\grad}[1]{\boldsymbol{\nabla}#1}
\newcommand{\diverg}[1]{\boldsymbol{\nabla}\cdot #1}

\newcommand{\Om}{\vect{\Omega}_\text{m}}

\newcommand{\Imantle}{\vect{I}_\text{m}}
\newcommand{\Icore}{\vect{I}_\text{c}}
\newcommand{\vw}{\textbf{w}}
\newcommand{\vz}{\hat{\vect{z}}}

\submitjournal{PSJ}
\received{April 1 2020}
\revised{May 13 2020}
\accepted{May 14 2020}

\begin{document}

\title{Inertial modes of a freely rotating ellipsoidal planet and their relation to nutations}

\correspondingauthor{Jeremy Rekier}
\email{jeremy.rekier@observatory.be}                        

\author[0000-0003-3151-6969]{Jeremy Rekier}
\affiliation{Royal Observatory of Belgium \\
 3 avenue circulaire \\
 Brussels, 1180, Belgium}

\author[0000-0002-7679-3962]{Santiago A. Triana}
\affiliation{Royal Observatory of Belgium \\
 3 avenue circulaire \\
 Brussels, 1180, Belgium}

\author[0000-0001-6732-4104]{Antony Trinh}
\affiliation{Lunar and Planetary Lab\\
 University of Arizona\\
 Tucson, AZ 85721-0092}

\author[0000-0002-9516-8572]{V\'eronique Dehant}
\affiliation{Royal Observatory of Belgium \\
 3 avenue circulaire \\
 Brussels, 1180, Belgium}

\begin{abstract}
We compute the inertial modes of a freely rotating two-layer planetary model with an ellipsoidal inviscid fluid core and a perfectly rigid mantle. We present a method to derive analytical formulae for the frequencies of the \emph{Free Core Nutation} (FCN) and \emph{Chandler Wobble} (CW) which are valid to all orders of the dynamical flattening of the core and mantle, and we show how the FCN and CW are the direct generalisation of the purely fluid \emph{Spin-Over mode} (SO) and of the \emph{Eulerian Wobble} (EW) to the case where the mantle can oscillate freely around a state of steady rotation. Through a numerical computation for an axisymmetric (oblate spheroidal) planet, we demonstrate that all other inertial modes of the steadily rotating fluid core are also free modes of the freely rotating two-layer planet.
\end{abstract}

\section{Introduction}

The rotation of a planet depends on its internal structure via its moments of inertia. In particular, the misalignment between the rotation axis of a planet and its moment of inertia axis, and the misalignment between the rotation axis of a planet and of its liquid core gives rise to two free modes of rotation known as the \emph{Chandler Wobble} (CW) and the \emph{Free Core Nutation} (FCN), respectively. In the planetary reference frame, the CW, if excited, is a long period prograde rotation of the instantaneous axis of rotation around the principal axis of largest moment of inertia of the planet, while the FCN, if exited, is a retrograde motion of nearly diurnal frequency around the same axis.\footnote{This motion is sometimes referred to as the \emph{Nearly Diurnal Free Wobble} (NDFW) in the planetary frame of reference.} When the gravitational pull on the planet from the sun, the moon and other planets is taken into account, these two modes can resonantly amplify the planet's forced nutation. Concerning the Earth, nutations are observed by using Very Long Baseline Interferometry (VLBI) and modeled precisely, accounting for deformation and coupling mechanisms at the core-mantle boundary such as pressure and electromagnetic couplings. A solid deformable inner core inside the liquid core is also accounted for (see e.g. \citet{dehant2015}). Concerning the planet Mars that possesses precession and nutations as well, the joint efforts of the current RISE \citep{Folkner2018} and future LaRa \citep{Dehant2020} lander missions aim to detect such amplification in the nutation signal of Mars measured directly from its surface. Such a detection would allow an estimate of Mars's FCN frequency, from which Mars' polar moment of inertia can be deduced; this can then be used to build radial profiles of the density structure of the planet, and in particular constrain its core size \citep{Dehant2020}. Usually, such a computation is based on the formalism of \citet{Sasao1980} in which the rotational dynamics of the fluid core is described in terms of its angular momentum which is assumed to be proportional to a mean flow resembling a solid-body rotation (see Sec.~\ref{sec:IT}). This has a couple of important advantages like that of being easily adaptable to accommodate the existence of a non-rigid (elastic) mantle. However, as we show in the present work, it is, by design, valid only to first order in the planet's dynamical flattening parameters. This casts doubts on the results of previous studies attempting to push the formalism of \citet{Sasao1980} beyond first order to evaluate the effects of trixiality \citep{VanHoolst2002,Chen2010,dehant2015,Guo2019,Shen2019}. 

In addition to the FCN and CW, the liquid core of a rotating star or planet is known to support the existence of oscillations known as \emph{inertial modes} caused by the restoring effect of the Coriolis force. Experimental studies in laboratory have shown that these inertial modes  can be excited by the action of tidal deformation at the fluid boundary, see \emph{e.g.} \citet{Malkus1968,Aldridge1969,Kelley2007} and \citet{Morize2010}. However, the role of the inertial modes on the rotation of planets and other astrophysical objects remains unclear. The main reason for this being that most studies, so far, have focused on the response of the fluid core in situations where the rotation of the planet is prescribed, \emph{i.e.} known \emph{a priori}, therefore disregarding the action of the fluid core on the rotation via the forces acting at the \emph{Core Mantle Boundary} (CMB), see \citet{LeBars2015} for a review. In a previous work, we included the action of the fluid core on the free rotation of a two-layer planet with a viscous liquid core \citep{Triana2019}. We showed that the FCN and CW naturally emerge among the spectrum of inertial modes and that complex interactions between the eigenmodes take place when the moment of inertia of the mantle varies. Due to the difficulty in dealing with both the presence of a thin viscous boundary layer and the ellipticity of the CMB, we could not explore the regime of very low viscosity relevant to planetary and astrophysical objects. 

In the present work, we complement the analysis of \citeauthor{Triana2019}, focusing specifically on the effects of the pressure torque between the fluid core and the solid mantle. We do so by computing the free modes of rotation of a two-layer ellipsoidal planet model with a perfectly rigid mantle and an inviscid fluid core. When one sets the viscosity of the fluid to zero, the viscous boundary layer at the CMB and the various \emph{shear layers} that it generates inside the core disappear and the flow pattern inside the core becomes smooth. This enables us to derive the free modes of rotation at all values of the core and mantle's ellipticities.

Previous studies have attempted to mathematically model the rotational modes of a terrestrial planet based on the displacement field approach whereby the motion of the core and mantle are computed using the set of gravito-elastic equations which follow from the local conservation of momentum \citep{Smith1974,Smith1977,Rogister2001,Rochester2014,Seyed-Mahmoud2017}. However, these are limited to considering planetary models that are only slightly aspherical. There is also some doubt regarding the ability of these work to capture adequately the effects of the presence of a solid inner core which is known to generate singularities within the displacement field \citep{Rieutord2000}.

This work is structured as follows. In Sec.~\ref{sec:model}, we present the two-layer planet model and its equations of motion. We then consider the special case of a core flow with a uniform vorticity and derive analytical formulae for the FCN and CW frequencies. In Sec.~\ref{sec:results}, we compute the inertial modes of a two-layer planet with an axisymmetric CMB numerically using the set of oblate spheroidal coordinates and compare the value of the FCN frequency to the analytical expressions of \citet{Sasao1980} and our own formalism. Results are discussed in Sec.~\ref{sec:discussion}.

\section{Model}
\label{sec:model}

\begin{figure}
\center
\includegraphics[width=0.6\textwidth]{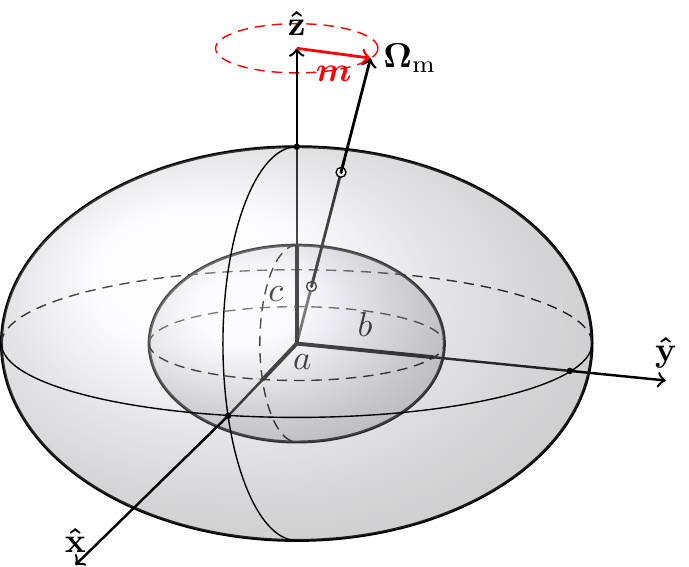}
\caption{The two-layer planet model with an ellipsoidal core and surface. The cartesian basis $\{\vect{\hat{x}},\vect{\hat{y}},\vect{\hat{z}}\}$ is aligned with the principal axes of inertia of the core and the rigid mantle. $\vect{\Omega}_m$ is the rotation vector of the mantle relative to the inertial frame of reference and is related to the polar figure axis, $\vect{\hat{z}}$, by Eq.~(\ref{eq:Omz}), where $|\vect{m}|\ll1$.}
\label{fig:model}
\end{figure}

Fig.~\ref{fig:model} represents the planetary model used throughout this work. The usual way to compute the rotation of this two-layer model is via the system of equations expressing the conservation of angular momentum for each layer. In the reference frame attached to the mantle rotating at angular velocity $\Om$ relative to the inertial frame, and in the absence of external forces, these read:
\begin{align}
\partial_t\vect{L}_\text{m}+\Om\times\vect{L}_\text{m}&=\vect{\Gamma}~,\label{eq:dtLmantle}\\
\partial_t\vect{L}_\text{c}+\Om\times\vect{L}_\text{c}&=-\vect{\Gamma}~,\label{eq:dtLcore}
\end{align}
where $\vect{L}_\text{c}$ and $\vect{L}_\text{m}$ denote the angular momentum of the core and mantle respectively and $\vect{\Gamma}$ represents the total torque that the fluid core exerts on the mantle. All these quantity are measured relative to the inertial frame. In the absence of other external torques, this must be balanced by the torque from the core on the mantle. The addition of Eqs.~(\ref{eq:dtLmantle}) and (\ref{eq:dtLcore}) then yields:
\begin{equation}
\partial_t\vect{L}+\Om\times\vect{L}=\vect{0}~,\label{eq:dtL}
\end{equation}
where $\vect{L}\equiv\vect{L}_\text{c}+\vect{L}_\text{m}$ is the total angular momentum of the planet. Traditionally, one chooses to solve Eqs.~(\ref{eq:dtLcore}) and (\ref{eq:dtL}). The difficulty then lies in the determination of $\vect{\Gamma}$ which expresses the coupling between the core and the mantle. For the case of a perfect fluid with zero viscosity and in the absence of magnetic field the only contribution comes from the pressure forces at the CMB which give rise to the pressure torque:
\begin{equation}
\vect{\Gamma}=\oint_\text{CMB}P\vect{r}\times\vect{\hat{n}}
\label{eq:GammaP}
\end{equation}
where $P$ is the pressure inside the fluid and $\vect{r}$ is the position vector. The integration runs over the whole surface of the CMB and $\vect{\hat{n}}$ denotes the radially-outward normal vector to this surface. In principle, one should add the increment of gravity resulting from the change in the repartition of mass to Eq.~(\ref{eq:GammaP}). In our simple model, however, this contribution is zero as the distribution of masses is kept constant due to the perfect rigidity of the mantle and the incompressibility of the fluid core. \citet{Sasao1980} have shown how to avoid the computation of the pressure torque explicitly by replacing Eq.~(\ref{eq:dtLcore}) with an equation involving the mean rotation of the fluid. This approach, however, is only valid in the limit of small flattening of the CMB (see Sec.~\ref{sec:IT}). In the present work, we use a different approach based on the equation giving the evolution of the flow inside the fluid core. In the special case where the flow vorticity is spatially uniform, the problem reduces to determining the evolution of the components of the vorticity in time. We can do so analytically by working from the equation of vorticity (the curl of the momentum equation) in a way that is formally similar to the formalism of \citeauthor{Sasao1980} but which has the advantage to be valid for all values of the flattening.

In the remainder of this section, we present the equations governing the dynamics of the fluid core coupled to planetary rotation. We then present their analytical resolution in the special case of a flow with uniform vorticity and provide a comparison to the formalism of \citeauthor{Sasao1980}. 

\subsection{Equations for a general flow}

We focus on the small departures of the flow from a solid-body rotation and write:
\begin{equation}
\vect{v}=\Om\times\vect{r}+\vect{u}~,
\label{eq:vu}
\end{equation}
where $\vect{v}$ and $\vect{u}$ denote the velocity field measured in the inertial frame and the mantle frame, respectively. Assuming that the fluid is incompressible and has zero viscosity, the equation of motion in the mantle frame and to first order in $\vect{u}$ reads \citep{Triana2019}:
\begin{equation}
\partial_t\vect{u}+2\Om\times\vect{u}+\partial_t\Om\times\vect{r}=-\grad{p}~,
\label{eq:momentum}
\end{equation}
where $\rho$ denotes the fluid density and we have identified the \emph{reduced pressure}, $p$, as:
\begin{equation}
p\equiv\frac{P}{\rho}-\frac{1}{2}|\vect{\Omega}_m\times\vect{r}|^2~,
\label{eq:reducedp}
\end{equation}
where the second term represents the centrifugal potential which, in the absence of external forces, is the only contribution to the total potential. The flow must satisfy the \emph{no-penetration condition} at the CMB:
\begin{equation}
\left.\vect{u}\cdot\vect{\hat{n}}\right|_\text{CMB}=0~.
\label{eq:u.n=0}
\end{equation}
In this work, we focus on the small oscillations of the planet about a steady rotation whereby $\Om$ can be decomposed as:
\begin{equation}
\Om=\Omega_0(\vect{\hat{z}}+\vect{m})~,
\label{eq:Omz}
\end{equation}
where the cartesian basis vector, $\vect{\hat{z}}$, is chosen as the mean axis of rotation (see Fig.~\ref{fig:model}) and we have $|\vect{m}|\ll1$. In what follows, we set $\Omega_0=1$ for simplicity which amounts to measure frequencies in cycles per day.
As we are looking for solutions that are oscillatory, we use the following Fourier decomposition:
\begin{align}
\vect{u}&=\vect{u}~e^{i\omega t}+\vect{u}^*e^{-i\omega t}~,\label{eq:uiwt}\\
p&=p~e^{i\omega t}+p^*e^{-i\omega t}~,\label{eq:piwt}\\
\vect{m}&=\vect{m}~e^{i\omega t}+\vect{m}^*e^{-i\omega t}~,\label{eq:miwt}
\end{align}
where a $*$ denotes the complex conjugation and where we have used the same symbol to denote a quantity and its Fourier component at frequency $+\omega$ for simplicity. Inserting Eq.~(\ref{eq:Omz}) and Eqs.~(\ref{eq:uiwt}) to (\ref{eq:miwt}) into Eq.~(\ref{eq:momentum}), keeping terms up to first order in $\vect{m}$ and $\vect{u}$ yields:
\begin{align}
i\omega\vect{u}+2\vect{\hat{z}}\times\vect{u}+i\omega(\vect{m}\times\vect{r})&=-\grad{p}~,\label{eq:momentumomega}\\
-i\omega\vect{u}^*+2\vect{\hat{z}}\times\vect{u}^*-i\omega(\vect{m}^*\times\vect{r})&=-\grad{p}^*~.
\end{align}
From Eq.~(\ref{eq:momentumomega}), we can isolate $\vect{u}$ and express it in terms of in terms of $p$ and $\vect{m}$ only. Upon using the incompressibility condition ($\diverg{\vect{u}}=0$), and the condition of no-penetration Eq.~(\ref{eq:u.n=0}), we then arrive respectively to (see Appendix~\ref{sec:inertialmodesspheroid}):
\begin{equation}
-\omega^2\vect{\nabla}^2{p}+4(\vect{\hat{z}}\cdot\vect{\nabla})^2p=-4(\vect{\hat{z}}\cdot\vect{m})~,
\label{eq:poincareeq}
\end{equation}
which is valid inside the whole core and
\begin{multline}
-\omega^2\vect{\hat{n}}\cdot(\grad{p}+i\omega(\vect{m}\times\vect{r}))+2i\omega(\vect{\hat{z}}\times\vect{\hat{n}})\cdot(\grad{p}+i\omega(\vect{m}\times\vect{r}))+4(\vect{\hat{z}}\cdot\vect{\hat{n}})\vect{\hat{z}}\cdot(\grad{p}+i\omega(\vect{m}\times\vect{r}))|_\text{CMB}=0~.
\label{eq:poincareeqbc}
\end{multline}
which must be satisfied at the CMB. When the mantle is in steady rotation, one has $\vect{m}=\vect{0}$. Eq.~(\ref{eq:poincareeq}) then reduces to the well-known \emph{Poincar\'e equation}, the solutions of which are global oscillations of the flow called \emph{inertial modes} and their frequencies satisfy $-2<\omega<2$ (see \emph{e.g.} Sec. 8.3.3 of \citet{rieutord2015}). In his classic paper, \citet{Bryan1889} gave the general analytical implicit expression of the inertial modes frequencies for an axisymmetric ellipsoid (those are reproduced in Appendix \ref{sec:inertialmodesspheroid}). His work was later extended by that of \citet{Hough1895} to the geometry of the triaxial ellipsoid. In both cases, the pressure field associated to each mode can be represented as a polynomial function in the cartesian coordinates \citep{kudlick1966,Zhang2001,vantieghem2014}, see also \citet{Rekier2018}. The inertial mode of lowest degree is called the \emph{Spin-Over mode} (SO) and its flow, when it is observed from the mantle frame, resembles a solid-body rotation around an axis perpendicular to the rotation vector of the mantle. It is the only inertial mode that carries a net amount of total angular momentum when integrated over the whole core and is therefore especially important for rotation as we show later.  

In order to obtain the evolution of $\vect{m}$, we must use Eq.~(\ref{eq:dtLmantle}) and express the angular momentum of the mantle in terms of its tensor of inertia, $\vect{I}_\text{m}$, and its angular velocity:
\begin{equation}
\vect{L}_\text{m}=\vect{I}_\text{m}\cdot\vect{\Omega}_m~.
\end{equation}
To obtain the right-hand side of Eq.~(\ref{eq:dtLmantle}), we must write the pressure torque Eq.~(\ref{eq:GammaP}) in terms of the reduced pressure, $p,$ and the increment of angular velocity $\vect{m}$. From the definition Eq.~(\ref{eq:reducedp}), one has, to first order in $\vect{m}$:
\begin{equation}
P=\rho\left(p+\frac{1}{2}|\vect{\Omega}_m\times\vect{r}|^2\right)\approx \rho \; p+\frac{\rho}{2}|\vect{\hat{z}}\times\vect{r}|^2+\rho(\vect{\hat{z}}\times\vect{r})\cdot(\vect{m}\times\vect{r})~.
\label{eq:Pp}
\end{equation}
In Sec.~\ref{sec:results}, we solve Eqs.~(\ref{eq:poincareeq}), (\ref{eq:poincareeqbc}) and Eq.~(\ref{eq:dtLmantle}) numerically for an axisymmetric oblate ellipsoidal fluid core using a set of oblate spheroidal coordinates, in a manner similar to what we presented in \citet{Rekier2018}. In the remainder of the present section, we focus on the special case where the flow inside the core has a uniform vorticity for which the free modes of the two-layer planet can be computed analytically.

\subsection{Flow with a uniform vorticity}
\label{sec:uniformw}

In his classic paper, \citet{Poincare1885} looked for solutions of the fluid flow in an ellipsoid characterised by a spatially uniform (yet possibly time-dependent) vorticity. Denoting this vorticity by $\vect{\nabla}\times\vect{u}=2\vect{w}$, the velocity field, $\vect{u}$, writes as:
\begin{equation}
\vect{u}=(\vect{w}\times\vect{r})+\grad{\psi}~,
\label{eq:upoincare}
\end{equation}
where $\psi$ is explicitly expressed in terms of $\vect{w}$ and the ellipsoid dimensions ($a$, $b$, $c$). This scalar function is there to adjust the flow to the no-penetration boundary condition Eq.~(\ref{eq:u.n=0}) and is zero for a spherical core. The explicit expressions of $\vect{u}$ and $\psi$ are given in Appendix~\ref{sec:poincaréflow}. We obtain the equation giving the time evolution of the vorticity by taking the curl of Eq.~(\ref{eq:momentum}):
\begin{equation}
\partial_t\vect{w}-\Om\cdot\grad{\vect{u}}+\partial_t\Om=\vect{0}~.
\label{eq:vorticity}
\end{equation}
This has the advantage to make the gradient of pressure vanish and the whole influence of the mantle on the fluid core is accounted for entirely via the variations of $\Om$. Under the assumption of uniform vorticity, the computation of the fluid flow inside the core reduces to the resolution of Eq.~(\ref{eq:vorticity}) for the three cartesian components of $\vect{w}$. In order to obtain the coupled motion of the core and mantle, we must add another equation giving the evolution of $\Om$. We choose to use Eq.~(\ref{eq:dtL}) as it does not require any explicit computation of the torque at the CMB. It does, however, require to evaluate the angular momentum of the whole planet relative to the inertial frame, $\vect{L}$. Using Eq.~(\ref{eq:vu}), this writes:
\begin{equation}
\vect{L}=(\Icore+\Imantle)\cdot\Om+\int_\text{core}\rho(\vect{r}\times\vect{u})~,
\label{eq:totalL}
\end{equation}
where the first term is proportional to the total tensor of inertia of the whole planet and the integral runs over the whole volume of the core. As the mantle is taken to be perfectly rigid, we can always chose the coordinates of the mantle frame to be aligned with the principale axes of inertia of the core and mantle at all time and parametrise the tensor of inertia as:
\begin{equation}
\underbrace{\begin{pmatrix} \text{A} & 0 & 0\\0 & \text{B} & 0\\0 & 0 & \text{C}\end{pmatrix}}_{\vect{I}}=\underbrace{\begin{pmatrix} \text{A}_f & 0 & 0\\0 & \text{B}_f & 0\\0 & 0 & \text{C}_f\end{pmatrix}}_{\vect{I}_\text{c}}+\underbrace{\begin{pmatrix} \text{A}_m & 0 & 0\\0 & \text{B}_m & 0\\0 & 0 & \text{C}_m\end{pmatrix}}_{\vect{I}_\text{m}}~.
\label{eq:InertiaTensor}
\end{equation}
In Appendix~\ref{sec:poincaréflow}, we give the explicit expressions of the components of the core angular momentum. In what follows, we parametrise the moments of inertia in terms of the following \emph{dynamical flattening} parameters (see p.154 of \citet{dehant2015}):
\begin{align}
\alpha_f&\equiv\frac{\text{C}_f-\frac{\text{A}_f+\text{B}_f}{2}}{\frac{\text{A}_f+\text{B}_f}{2}}~,&\beta_f&\equiv\frac{\text{B}_f-\text{A}_f}{\text{B}_f+\text{A}_f}~,\label{eq:alphafbetaf}
\end{align}
for the fluid core and 
\begin{align}
\alpha&\equiv\frac{\text{C}-\frac{\text{A}+\text{B}}{2}}{\frac{\text{A}+\text{B}}{2}}~,&\beta&\equiv\frac{\text{B}-\text{A}}{\text{B}+\text{A}}~,\label{eq:alphabeta}
\end{align}
for the whole planet.

As we are interested in the small oscillations of the flow around the axis of rotation, we write, in analogy with Eq~(\ref{eq:miwt}) and consistent with the notation of \citet{Sasao1980}:
\begin{equation}
\vect{w}=(\vect{m}_f~e^{i\omega t}+\vect{m}_f^*e^{-i\omega t})~,
\end{equation}
and we assume $|\vect{m}_f|\ll1$. The system of Eqs.~(\ref{eq:dtL}) and (\ref{eq:vorticity}) in the first order in $\vect{m}$ and $\vect{m}_f$ can be put in the following matrix form from which the components of these vectors in the $z$-direction decouple completely:
\begin{equation}
\begin{scriptsize}
\underbrace{\left(
\begin{array}{cccc}
 \omega  & -\frac{i (\alpha -\beta )}{(1-\beta)} & \frac{\text{A}_f
   \left(1-\alpha _f\right) \left(1+\alpha _f-2 \beta _f\right)}{\text{A} \left(1-\beta
   _f\right){}^2}~\omega  & \frac{i \text{A}_f \left(1-\alpha _f\right) \left(1+\alpha
   _f+2 \beta _f\right)}{\text{A} \left(1-\beta _f^2\right)} \\
 \frac{i (\alpha +\beta )}{(1+\beta)} & \omega  & -\frac{i \text{A}_f
   (1-\beta) \left(1-\alpha _f\right) \left(1+\alpha _f-2 \beta _f\right)}{\text{A} (1+\beta)
   \left(1-\beta _f\right){}^2} & \frac{\text{A}_f (1-\beta)  \left(1-\alpha
   _f\right) \left(1+\alpha _f+2 \beta _f\right)}{\text{A} (1+\beta) \left(1-\beta
   _f^2\right)}~\omega  \\
 \omega  & 0 & \omega  & \frac{i \left(1+\alpha _f+2
   \beta _f\right)}{(1+\beta _f)} \\
 0 & \omega  & -\frac{i \left(1+\alpha _f-2 \beta _f\right)}{(1-\beta
   _f)} & \omega  \\
\end{array}
\right)}_{M}
\end{scriptsize}
\left(
\begin{array}{c}
m^x\\m^y\\m_f^x\\m_f^y
\end{array}
\right)=0~
\label{eq:Mmxmy}
\end{equation}
The problem of finding the free modes frequencies $\omega$, then reduces to solving the polynomial equation which results from imposing:
\begin{equation}
\text{det}[M]=0~.
\label{eq:detM}
\end{equation}
This polynomial equation is in fact \emph{biquadratic} in the frequency $\omega$ and so its roots come in pairs that have the same magnitude and opposite signs. The set of individual modes associated to each root can be classified into prograde and retrograde motions (relative to the mean planetary rotation). In the end, there are only two independent physical motions as a prograde (respectively retrograde) mode of frequency $\omega<0$ is equivalent to a retrograde (respectively prograde) mode of $\omega>0$. Here we follow the convention of \citet{dehant2015} and others and represent the retrograde modes using negative frequencies.
The explicit expressions of the two independent frequencies are too lengthy to reproduce here. Instead, we provide the expressions of their Taylor expansions in the dynamical flattening parameters:
\begin{align}
\omega=-1-&\frac{\text{A} \alpha _f}{\text{A}_m}+\frac{\text{A} \left(2 (\alpha +\beta ) \left(\text{A}-\text{A}_m\right) \alpha
   _f-2 \left(\text{A}-\text{A}_m\right) \alpha _f \beta _f-\left(\text{A}-4 \text{A}_m\right) \beta _f^2\right)}{2 \text{A}_m^2}+\dots~,\label{eq:fcnseriestri}\\
\omega&=\frac{\text{A} \sqrt{|\alpha ^2-\beta ^2|}}{\text{A}_m}-\frac{\text{A} \sqrt{|\alpha ^2-\beta ^2|} \left(\text{A}-\text{A}_m\right)
   \left(\beta +\alpha _f-\beta _f\right)}{\text{A}_m^2}+\dots~.\label{eq:cwseriestri}
\end{align}
We identify these frequencies to those of the FCN and CW, respectively. The former presents a nearly diurnal retrograde nutation as measure in the planetary frame of reference. The latter is a prograde nutation of small frequency (and thus long period) in that same frame.

In anticipation of the next section, we now turn to the special case where the planet and its liquid core are axisymmetric around the mean rotation axis, $\vect{\hat{z}}$. In such case, $\beta=\beta_f=0$ and Eq.~(\ref{eq:Mmxmy}) can be further simplified in terms of the following variables:
\begin{align}
m^+&=\frac{1}{\sqrt{2}}(-m^x+i m^y)~,&m^-&=\frac{1}{\sqrt{2}}(m^x+i m^y)~,
\end{align}
and analogous expressions for the components of $\vect{m}_f$. These new variables are the components of $\vect{m}$ in the cartesian canonical basis (see \emph{e.g.} p.~193 of \citet{Trinh2019thesis}). The equations for $m^+$ and $m_f^+$ on the one-hand and for $m^-$ and $m_f^-$ on the other hand decouple completely so that we end up with:
\begin{equation}
\left(
\begin{array}{cc}
 \omega-\alpha & \frac{\text{A}_f \left(1-\alpha _f^2\right)}{\text{A}}\left(1+\omega\right) \\
 \omega & \omega+(1+\alpha _f) \\
\end{array}
\right)
\left(
\begin{array}{c}
m^-\\m_f^-
\end{array}
\right)=0~.
\label{eq:Mm+m-}
\end{equation}
Setting the determinant of the above matrix to zero gives the following two independent frequencies:
\begin{footnotesize}
\begin{equation}
\omega=-\frac{\text{A}_f \left(1-\alpha_f^2\right)-\text{A} \left(1-\alpha+\alpha_f\right)\pm\sqrt{4 \text{A} \alpha \left(1+\alpha_f\right) \left(\text{A}-\text{A}_f \left(1-\alpha_f^2\right)\right)+\left(\text{A}
   \left(1-\alpha+\alpha_f\right)-\text{A}_f \left(1-\alpha_f^2\right)\right){}^2}}{2\left(\text{A}-\text{A}_f \left(1-\alpha_f^2\right)\right)}~.
\label{eq:omegam+m-}
\end{equation}
\end{footnotesize}
By taking the series expansion of Eq.~(\ref{eq:omegam+m-}) in $\alpha$ and $\alpha_f$, assuming that these two parameters are small quantities of the same order, one recovers more familiar expressions for the FCN and the CW in the axisymmetric case:
\begin{align}
\omega=-1&-\frac{\text{A} \alpha_f}{\text{A}_m}+\frac{\text{A} \alpha \left(\text{A}-\text{A}_m\right) \alpha_f}{\text{A}_m^2}-\frac{\text{A} \left(\text{A}-\text{A}_m\right) \alpha_f \left(-\text{A}_m \alpha_f^2+\text{A} \alpha
   \left(\alpha+\alpha_f\right)\right)}{\text{A}_m^3}+\dots~,\label{eq:fcnseries}\\
\omega=&\frac{\text{A} \alpha}{\text{A}_m}-\frac{\text{A} \alpha \left(\text{A}-\text{A}_m\right) \alpha_f}{\text{A}_m^2}+\frac{\text{A}\left(\text{A}-\text{A}_m\right)\alpha \left(-\text{A}_m \alpha_f^2+\text{A} \alpha_f 
   \left(\alpha+\alpha_f\right)\right)}{\text{A}_m^3}+\dots~.\label{eq:cwseries}
\end{align}
From Eq.~(\ref{eq:fcnseries}), one sees that the FCN has an exactly diurnal frequency in the case where the fluid core is spherical ($\alpha_f$). This is true to all orders in $\alpha$ and $\alpha_f$ as can be shown by working directly from Eq.~(\ref{eq:omegam+m-}). From Eq.~(\ref{eq:cwseries}), we see that the frequency of the CW remains proportional to the dynamical flattening of the whole planet, $\alpha$, in that same limit.
In Sec.~\ref{sec:results}, we demonstrate the validity of Eq.~(\ref{eq:omegam+m-}) by comparing it to the result of a direct numerical integration of Eq.~(\ref{eq:poincareeq}). We also provide a comparison to the formula obtained using the formalism of \citet{Sasao1980} to which we now turn.

\subsection{The inertial torque approximation}
\label{sec:IT}
Rather than working directly from the vorticity Eq.~(\ref{eq:vorticity}), most studies prefer to compute the motion of the fluid core from the conservation of angular momentum. In theory, this amounts to solve Eq.~(\ref{eq:dtLcore}) which requires the computation of the pressure torque contribution Eq.~(\ref{eq:GammaP}) to $\vect{\Gamma}$. In practice, however, \citet{Sasao1980} have shown how Eq.~(\ref{eq:dtLcore}) can be replaced by the following:
\begin{equation}
\partial_t\vect{L}_\text{c}-\vect{w}_\text{T}\times\vect{L}_\text{c}\approx\mathcal{O}(\epsilon^2)~,
\label{eq:dtLtisserand}
\end{equation}
where $\vect{w}_\text{T}$ denotes the angular velocity of the \emph{Tisserand frame} of the fluid core, which is \emph{defined} via 
\begin{equation}
\vect{L}_\text{c}=\Icore\cdot(\Om+\vect{w}_\text{T})~.
\label{eq:tisserand}
\end{equation}
The right-hand side of Eq.~(\ref{eq:dtLtisserand}) only involves terms that are second order or more in the flattening parameters $\alpha_f$ and $\beta_f$ here assumed to be proportional to a single small parameter $\epsilon$ (see \citet{dehant2015}, p.~271 and \citet{Trinh2019thesis}, p.~79 for details). The usage of Eq.~(\ref{eq:dtLtisserand}) over Eq.~(\ref{eq:dtLcore}) is usually referred to as the \emph{inertial torque approximation}. From the cartesian components of the core angular moment, one can verify that, \emph{for all considerations of angular momentum}, $\vect{w}_\text{T}$ is equivalent to 
$\vect{w}$ to first order in the flattening so that the latter can be used in place of the former in Eq.~(\ref{eq:dtLtisserand}) (Appendix~\ref{sec:poincaréflow}). In this approximation, Eq.~(\ref{eq:Mmxmy}) gets replaced by:
\begin{equation}
\left(
\begin{array}{cccc}
 \omega  & -\frac{i (\alpha -\beta )}{(1-\beta)} & \frac{ 
   \text{A}_f}{\text{A}}~\omega  & \frac{i \text{A}_f \left(1+\beta _f\right)}{\text{A} \left(1-\beta _f\right)} \\
 \frac{i (\alpha +\beta )}{(1+\beta)} & \omega  & -\frac{i \text{A}_f(1-\beta)}{\text{A} (1+\beta)} & \frac{\text{A}_f (1-\beta)\left(1+\beta _f\right)}{\text{A} (1+\beta) \left(1-\beta _f\right)}~ \omega  \\
 \omega  & 0 & \omega  & \frac{i 
   \left(1+\alpha _f\right)}{(1-\beta _f)} \\
 0 & \omega   &
  -\frac{i  \left(1+\alpha _f\right)}{\left(1+\beta _f\right)} & \omega   \\
\end{array}
\right)\left(
\begin{array}{c}
m^x\\m^y\\m_f^x\\m_f^y
\end{array}
\right)=0~.
\label{eq:MITa}
\end{equation}
The frequencies, $\omega$, derived from Eq.~(\ref{eq:MITa}) are equivalent to those derived from Eq.~(\ref{eq:Mmxmy}) only to first order in the flattening parameters, $\alpha$, $\beta$, $\alpha_f$ and $\beta_f$. 

\subsection{The Spin-Over mode}
\label{sec:SOmode}

In Sec.~\ref{sec:uniformw}, we have seen how the FCN and CW wobble are natural oscillations of the two-layer system when the `wobbly' motion of the mantle is coupled to a core flow of uniform vorticity. From Eq.~(\ref{eq:Mmxmy}), we can illustrate the relation between the FCN and the SO by considering the special case where $\vect{m}=\vect{0}$. Physically, this corresponds to the situation where the mantle is forced to remain in a state of steady rotation at all time. Mathematically, this amounts to ignore the first two rows and columns of $M$ so that Eq.~(\ref{eq:Mmxmy}) reduces to:
\begin{equation}
\left(
\begin{array}{ccc}
 \omega  & \frac{i \left(1+\alpha _f+2\beta _f\right)}{(1+\beta _f)} \\
 -\frac{i \left(1+\alpha _f-2\beta _f\right)}{(1-\beta _f)} & \omega \\
\end{array}
\right)
\left(\begin{array}{c}
m_f^x\\m_f^y
\end{array}
\right)=0~,
\end{equation}
this gives a single independent frequency (using the same convention of sign as in Sec.~\ref{sec:uniformw}):
\begin{equation}
\omega=-\sqrt{\frac{(1+\alpha_f)^2-4\beta_f^2}{(1-\beta_f)(1+\beta_f)}}~.
\label{eq:omegaSO}
\end{equation}
Upon using Eqs.~(\ref{eq:alphafabc}) and (\ref{eq:betafabc}) expressing $\alpha_f$ and $\beta_f$ in terms of the dimensions of the ellipsoidal core ($a$, $b$, $c$), one arrives to the more familiar expression (see \emph{e.g.} Eq.~(3.21) of \citet{vantieghem2014} or Eq.~(A.23) of \citet{Rekier2018}):
\begin{equation}
\omega=-\frac{2ab}{\sqrt{(a^2+c^2)(b^2+c^2)}}~.
\end{equation}
Note that carrying the same exercise with Eq.~(\ref{eq:MITa}) gives a different, and incorrect, expression for the SO frequency which matches Eq.~(\ref{eq:omegaSO}) only up to first order in the flattening parameters.

We see that, strictly speaking, the SO is a mode of the two-layer system only in the limit where the rotation of the mantle is steady. When this is not the case, such as in the context of planetary nutations, the SO leaves its place to the FCN. The flow inside the core is similar for both modes, having a uniform vorticity. The main difference is that the FCN is a combined motion of the core \emph{and} mantle. When the core is axisymmetric ($\beta_f=0$), Eq.~(\ref{eq:omegaSO}) reduces to:
\begin{equation}
\omega=-(1+\alpha_f)~.
\label{eq:omegaSOaxi}
\end{equation}
In Sec.~\ref{sec:results}, we use Eq.~(\ref{eq:omegaSOaxi}) to evaluate the discrepancy between the frequencies of the SO and the FCN numerically as a function of the core flattening and we compare it to the analytical expression Eq.~(\ref{eq:omegam+m-}).

\section{Results}
\label{sec:results}

In this section, we present the results from the direct numerical integration of Eqs.~(\ref{eq:poincareeq}), (\ref{eq:poincareeqbc}) and Eq.~(\ref{eq:dtLmantle}) for an axisymmetric oblate ellipsoidal fluid core. This computation is based on spectral decomposition introduced in \citet{Rekier2018} where the reduced pressure field is developed onto the basis of spheroidal harmonics of the form:
\begin{equation}
p(\xi,\vartheta,\varphi)=\sum_{\ell=0}^L\sum_{m=-\ell}^{\ell}p_{\ell,m}(\xi)\text{Y}_\ell^m(\vartheta,\varphi)~,
\label{eq:pYlm}
\end{equation}
where $\{\xi,\vartheta,\varphi\}$ are the set of oblate spheroidal coordinates and $L$ the degree of truncation. \citeauthor{Rekier2018} have shown how to use these coordinates to compute the inertial modes of a steadily rotating axisymmetric ellipsoid numerically by using the expansion Eq.~(\ref{eq:pYlm}) and a decomposition of the $p_{\ell,m}$ over the basis of Chebyshev polynomials, $\text{T}_k(x)$:
\begin{equation}
p_{\ell,m}(\xi)=\sum_{k=0}^Np_{k,\ell,m}\text{T}_k(x)~,
\label{eq:chebyshev}
\end{equation}
where $N$ is the degree of truncation. $x=\xi/\xi_\text{CMB}$ denotes the normalised $\xi$ coordinate and $\xi_\text{CMB}$ is the value of $\xi$ at the CMB. With an appropriate choice of normalisation, $\xi_\text{CMB}$ is a constant that depends only on the polar flattening. It can be shown that the component of the pressure torque Eq.~(\ref{eq:GammaP}) in the $z$-direction vanishes in these coordinates, due to axial symmetry in the azimutal $\varphi$ coordinate. The other components in the cartesian canonical basis reduce to:
\begin{align}
\Gamma^+&=~~\frac{4\pi}{5}\frac{ia(a^2-c^2)}{\sqrt{3}}P_{2,+1}\left(\xi_\text{CMB}\right)~,\label{eq:Gamma+}\\
\Gamma^-&=-\frac{4\pi}{5}\frac{ia(a^2-c^2)}{\sqrt{3}}P_{2,-1}\left(\xi_\text{CMB}\right)~,\label{eq:Gamma-}
\end{align}
where $a$ and $c$ are the semi-major and semi-minor axes of the core respectively and the $P_{2,\pm1}\left(\xi_\text{CMB}\right)$ are the spheroidal harmonics components of the physical pressure field with $\ell=2$ and $m=\pm1$ at the CMB. This should not be confused with the components of the reduced pressure, $p\left(\xi_\text{CMB}\right)$, to which it is related via Eq.~(\ref{eq:Pp}) which gives:
\begin{equation}
P_{2,\pm1}\left(\xi_\text{CMB}\right)=\rho \left(p_{2,\pm1}\left(\xi_\text{CMB}\right)-\frac{ac}{\sqrt{3}}~m^{\pm}\right)~.
\label{eq:pPm}
\end{equation}
From Eqs.~(\ref{eq:Gamma+}) and (\ref{eq:Gamma-}), we see that the geometry of the CMB restricts the family of flows coupled to the planetary rotation to those with a non-vanishing $\ell=2$ and $m=\pm1$ pressure component at the CMB. In particular, this limits the analysis to the set of modes that are antisymmetric with respect to the equatorial plane. 

Eqs.~(\ref{eq:dtLmantle}), (\ref{eq:poincareeq}) and (\ref{eq:poincareeqbc}) form a closed system of equations in $p$ and $\vect{m}$ that we solve numerically as a (polynomial) eigenvalue problem in the frequency $\omega$. For definiteness, we arbitrarily assume that the two layers of the planet have the same density and that the semi-major axes of the fluid core and of the whole planet are in the $\frac{1}{2}$ ratio. We also set the exterior flattening to zero, $\alpha=0$, \emph{i.e.} the outside surface of the planet is spherical. This leaves $\alpha_f$ as the only free parameter. In Appendix~\ref{sec:inertiatensor}, we provide the detailed expressions of the different moments of inertia involved in this computation.

\begin{figure}
\center
\includegraphics[width=0.9\textwidth]{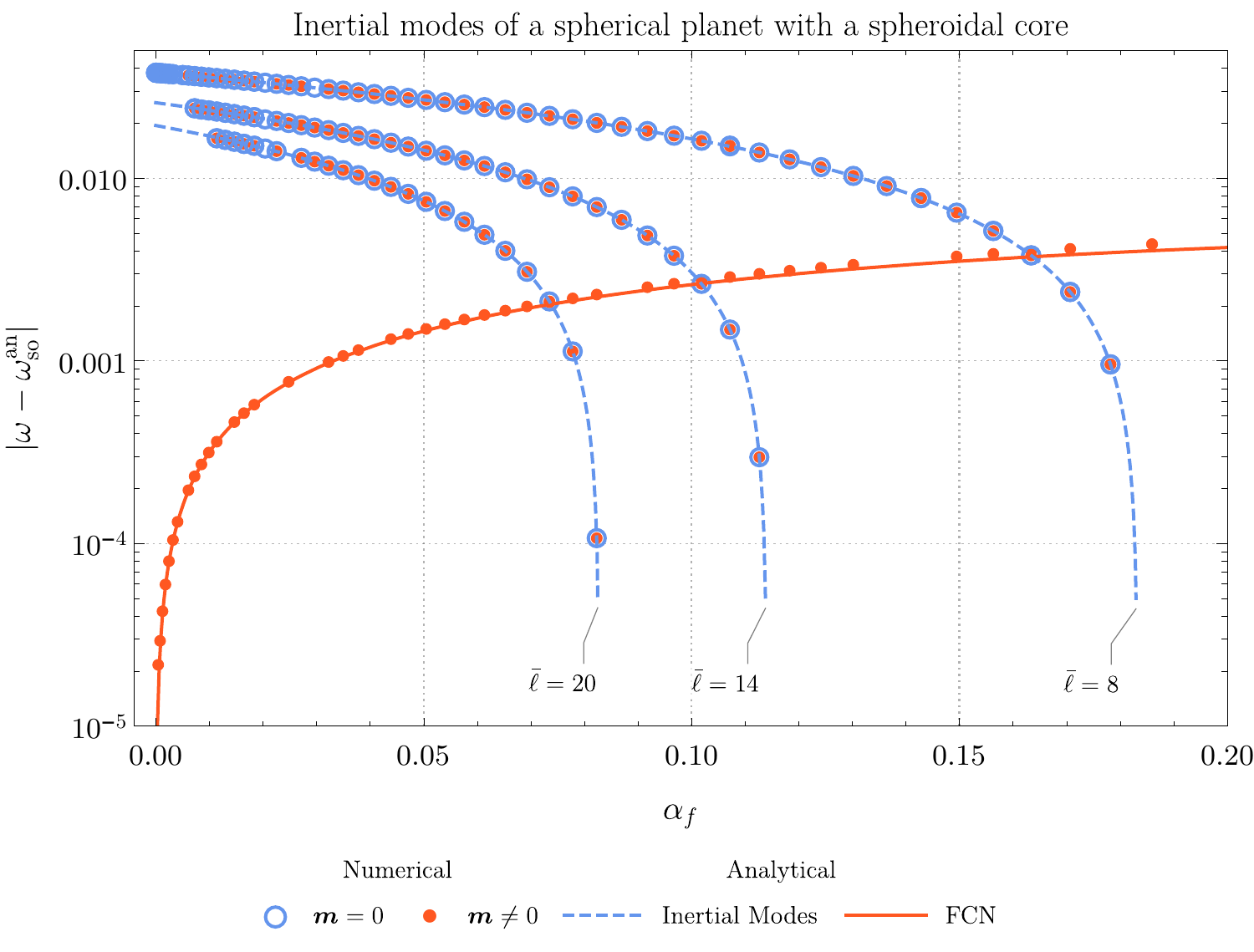}
\caption{Partial spectrum of eigenfrequencies of the two-layer planet as a function of the polar flattening of the fluid core. The blue circles are the frequencies of the inertial modes computed numerically for a planet in steady rotation ($\vect{m}=\vect{0}$). The red dots are the same thing for a planet with a non-steadily rotating (wobbly) mantle ($\vect{m}\neq\vect{0}$). The eigenvalues corresponding to the frequencies of inertial modes are the same in both cases except for the SO which gets replaced by the FCN. The blue dashed lines represent the analytical frequencies of the inertial modes of a steadily rotating planet (see Appendix~\ref{sec:inertialmodesspheroid}). These are labelled with the $\ell$-number of their highest degree component. The red curve represents the frequency of the FCN computed from Eq.~(\ref{eq:FCNalpha=0}). We have subtracted the value of the SO frequency Eq.~(\ref{eq:omegaSOaxi}) from all eigenvalues to accentuate the discrepancy with the FCN. All the frequencies in the figure correspond to modes with the azimuthal wave number $m=1$.}
\label{fig:wvwan}
\end{figure}

Fig.~\ref{fig:wvwan} represents the partial spectrum of eigenvalues of modes with the azimuthal wave number $m=1$ as a function of $\alpha_f$.
The blue circles are the inertial modes frequencies when the planet is in steady rotation ($\vect{m}=\vect{0}$). The red dots are the same thing for a planet with a non-steadily rotating (wobbly) mantle ($\vect{m}\neq\vect{0}$). The blue dashed lines represent the analytical frequencies of the inertial modes of a steadily rotating planet (see Appendix~\ref{sec:inertialmodesspheroid}). These are labelled with the $\ell$-number of their highest degree component. The frequencies are the same in both cases, the only exception being the SO which disappears from the spectrum when the planet has a non-steady rotation and is replaced by the FCN. In order to accentuate the discrepancy in the frequencies of these two modes, we have subtracted the SO frequency Eq.~(\ref{eq:omegaSOaxi}) from all the frequencies in the plot.
The red curve represents the frequency of the FCN computed from Eq.~(\ref{eq:omegam+m-}) which, for $\alpha=0$, simplifies to:
\begin{equation}
\omega=-1-\frac{\text{A}\alpha_f}{\text{A}_m+(\text{A}-\text{A}_m)\alpha_f^2}~.
\label{eq:FCNalpha=0}
\end{equation} 
It is in very good agreement with the values computed numerically even for large values of $\alpha_f$.
We observe that, the non-steady rotation of the mantle has no effect on the frequencies of the inertial modes other than the SO. This can be confirmed by looking at the ratio of the kinetic energy densities of the mantle and the core measured from the steadily rotating reference frame (see Appendix~\ref{sec:TOM}) as shown on Fig.~\ref{fig:Ekinr}.
\begin{figure}
\center
\includegraphics[width=0.8\textwidth]{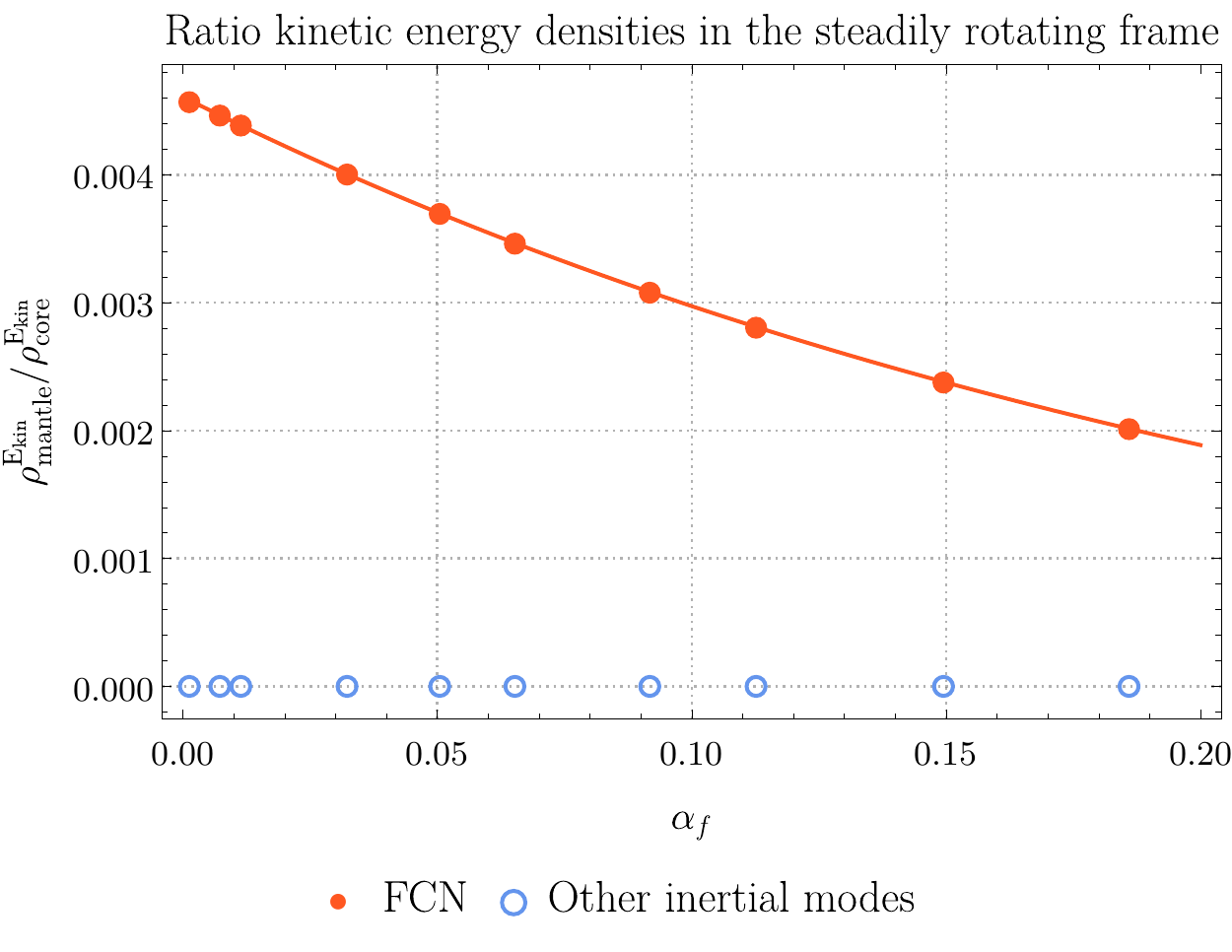}
\caption{Ratio of the kinetic energy densities of the mantle and the fluid core as measured from the steadily rotating frame. The red line corresponds to the analytical value computed for the solution of Eq.~(\ref{eq:Mm+m-}) with $\alpha=0$ and with a frequency given by Eq.~(\ref{eq:FCNalpha=0}).}
\label{fig:Ekinr}
\end{figure}
As we can see, the ratio is strictly zero for all modes other than the FCN. The red line corresponds to the analytical value computed from the resolution of Eq.~(\ref{eq:Mm+m-}) with $\alpha=0$ and with a frequency equal to Eq.~(\ref{eq:FCNalpha=0}). It is in good agreement with the numerical computation even for large values of the flattening. 

\begin{figure}
\center
\includegraphics[width=0.8\textwidth]{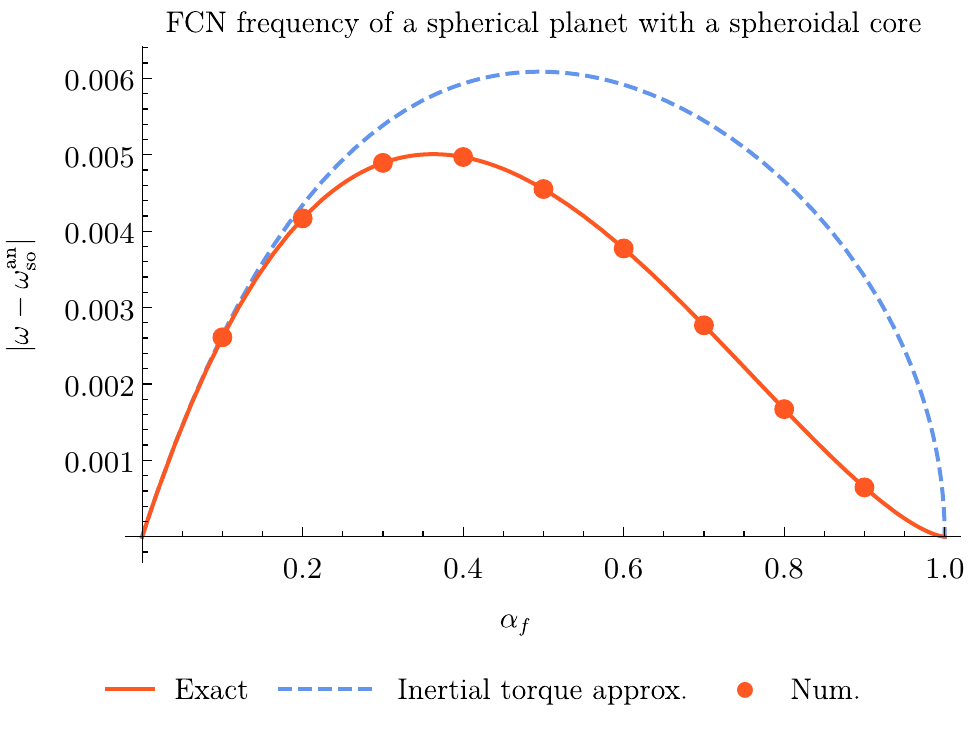}
\caption{Frequency of the FCN as a function of the core flattening. The red dots are the results from direct numerical integration and the red curve corresponds to the analytical formula of Eq.~(\ref{eq:FCNalpha=0}). The numerical and analytical results agree well for all values of $\alpha_f$. The blue dashed line represents the value of the frequency computed using the inertial torque approximation Eq.~(\ref{eq:FCNalpha=0IT}) which predicts the wrong value of the FCN frequency for large values of the flattening.}
\label{fig:wanvsnum}
\end{figure}

We can further illustrate the agreement between the numerical and analytical computations by looking at Fig.~\ref{fig:wanvsnum} which shows the same thing as Fig.~\ref{fig:wvwan} but focuses on the frequency of the FCN up to very large values of the core flattening, $\alpha_f\approx1$. 
The red curve represents the frequency obtained from Eq.~(\ref{eq:FCNalpha=0}) and the dots shows the numerical computation. The blue dashed line represents the value of the frequency computed using the inertial torque approximation described in Sec.~\ref{sec:IT} which, for $\alpha=0$, predicts:
\begin{equation}
\omega=-1-\frac{\text{A}}{\text{A}_m}\alpha_f~.
\label{eq:FCNalpha=0IT}
\end{equation}
We see that, in the example of the figure, the numerical computation starts to deviate markedly from Eq.~(\ref{eq:FCNalpha=0IT}) for $\alpha_f\gtrsim0.1$. The precise value of this threshold depends on the values of the moments of inertia, $\text{A}$ and $\text{A}_m$. We also see that Eq.~(\ref{eq:FCNalpha=0}) agrees well with the numerical computation for all values of  $\alpha_f$ thus proving the validity of the computation of Sec.~\ref{sec:uniformw} based on the vorticity equation.

\section{Discussion}
\label{sec:discussion}

We have presented a method to compute the free rotation of a simplified two-layer planet with an inviscid and incompressible fluid core and a perfectly rigid ellipsoidal mantle. This method is based on the resolution of the modified Poincar\'e Eq.~(\ref{eq:poincareeq}) coupled to the Liouville Eq.~(\ref{eq:dtLmantle}) which describes the `wobbling' motion of the mantle subjected to the pressure torque from the fluid at the CMB. We have found that the inertial modes of the fluid core enclosed in a steadily rotating mantle are also modes of the freely rotating two-layer planet. The only exception being the Spin-Over mode -- the simplest of the inertial modes -- which disappears from the spectrum and is replaced by the Free Core Nutation (Fig.~\ref{fig:wvwan}). The fact that this is the only mode that can influence the rotation of the planet (or be influenced by it) can be traced back to the fact that it is the only inertial mode that exerts a net pressure torque at the CMB, something that was already noted by \citet{Toomre1974}. This can be verified by looking at Eqs.~(\ref{eq:Gamma+}), (\ref{eq:Gamma-}) and (\ref{eq:pPm}) after realising that the reduced pressure components $p_{2,1}$ and $p_{2,-1}$ are zero at the CMB for all modes except the Spin-Over. This result demonstrated here for a freely rotating spheroid remains valid for a triaxial ellipsoid or when there are external tidal forces acting on the planet. In which case, the torque from the external potential must be added to the total torque Eq.~(\ref{eq:GammaP}) but it must also be included in the definition of the reduced pressure Eq.~(\ref{eq:reducedp}) so that nothing is changed.

Looking at Fig.~\ref{fig:wvwan} showing the evolution of the spectrum of eigenvalues as a function of the core flattening, we observe that the frequency of the FCN crosses path with that of the other inertial modes when $\alpha_f$ varies. In their study of the completeness of the set of inertial modes in a steadily-rotating ellipsoid, \citet{Backus2017} showed that the modes remain orthogonal even when such accidental degeneracies happen. In the same paper, the authors also demonstrated that the Poincar\'e Eq.~(\ref{eq:poincareeq}) and the boundary condition Eq.~(\ref{eq:poincareeqbc}) form a \emph{self-adjoint} system\footnote{For an introductory discussion on the `self-adjointness' of differential operators and its implication, see \emph{e.g} Sec.~10.1 and 10.2 of \citet{arfken2005}} when $\vect{m}=\vect{0}$, thus providing the mathematical explanation for the realness of the eigenfrequencies. Our work extends their conclusions to the case where the motion of the mantle is non-steady. \citet{Triana2019} have shown that the pictures drawn out from Fig.~\ref{fig:wvwan} changes significantly when the viscosity of the fluid is taken into account. The eigenvalues are then complex numbers and their imaginary parts represent the damping of the corresponding eigenmodes. Instead of crossing, the eigenvalues \emph{avoid} each other in the complex plane in a complicated manner that is not yet fully elucidated (see their Fig.~3 \& 4). A similar example of this phenomenon referred to as \emph{avoided crossing} was previously given by \citet{Rogister2009} who studied the influence of a thermally stratified liquid core on the rotational modes of the Earth. In both cases, avoided crossings result from the introduction of a non-adiabatic diffusion term in the dynamical equations (respectively viscous or thermal) which breaks their `self-adjointness'. \citet{birch2002} gave a similar interpretation in their study of the magneto-hydrodynamical equations for a one dimensional stratified medium where the avoided crossings result from the introduction of radiative cooling into the model. It should be noted that avoided crossings can also take place in systems that are self-adjoint. This is the case, for example, for the eigenfrequencies of the spheroidal oscillations of stars both in the adiabatic and non-adiabatic case. See \emph{e.g.} Chap.~5 of \citet{dalsgaard2014} or Sec.~11.3 of \citet{smeyers2010} for a review. See also \citet{Triantafyllou1991} for a more general description of the phenomenon and applications to problems in engineering.

We have given the exact analytical expressions of FCN and the CW frequencies of a triaxial two-layer planet. To second order in the flattening, these are given as Eqs.~(\ref{eq:fcnseriestri}) and (\ref{eq:cwseriestri}) which derive from the joint resolution of the Liouville equation for the mantle Eq.~(\ref{eq:dtLmantle}) and the equation of vorticity for the fluid core Eq.~(\ref{eq:vorticity}) based on the assumption that flow vorticity is uniform throughout the core. The task then reduces to solving Eq.~(\ref{eq:Mmxmy}). We have shown that the resulting formula for the FCN is valid to all orders in the flattening parameters, by comparing it to the numerical solution for a spheroidal core. This is illustrated on Fig.~\ref{fig:wanvsnum} where we have also plotted the same result based on the formalism of \citet{Sasao1980} summarised in Sec.~\ref{sec:IT}. In the same section, we have shown how that formalism, based on Eq.~(\ref{eq:dtLtisserand}), is only valid to first order in the flattening parameters.

Based on Eq.~(\ref{eq:Mmxmy}), valid for core flows of uniform vorticity, we have shown how the FCN reduces to the SO in the limit where the rotation of the mantle is steady. We can understand this by considering the fact that the pressure torque Eq.~(\ref{eq:GammaP}) vanishes identically when the CMB is spherical. This limit is equivalent to setting $a=c$ in Eqs.~(\ref{eq:Gamma+}) and (\ref{eq:Gamma-}) for an axi-symmetric core. The motion of the mantle then becomes independent to that of the fluid core and \emph{vice versa} and the motion of the planet is derived from Eq.~(\ref{eq:Mmxmy}) after setting $\vect{m}_f=\vect{0}$. The solution is that of fully rigid solid-body rotation with frequency:
\begin{equation}
\omega=\sqrt{\frac{\alpha^2-\beta^2}{1-\beta^2}}=\sqrt{\frac{(\text{C}-\text{A})(\text{C}-\text{B})}{\text{A}\text{B}}}~.
\end{equation}
We recognise the expression on the right-hand side as the frequency of the free \emph{Euler Wobble} (EW). If we imagine increasing the flattening of the CMB from zero, the core and mantle become coupled and the EW and SO disappear from the solutions to be replaced by the CW and the FCN. Both of which are coupled motions of the core and mantle, even though the CW is mostly a motion of the mantle and the FCN is, as its name indicates, mostly a motion of the fluid core.

In Sec.~\ref{sec:SOmode}, we have shown that, while the SO and FCN share a similar type of motion in the fluid core (one of uniform vorticity), it is improper to speak of the SO of a freely rotating planet. Some publications prefer to use the term \emph{Tilt-Over Mode} (TOM) \citep{Toomre1974,Noir2003,Cebron2010}, but this has the disadvantage to be ambiguous with yet another mode of the same name which results from the mismatch between the polar axis of a planet and its instantaneous rotation axis \citep{Dehant1996}. This mode has a purely diurnal frequency in the mantle frame, regardless of the planet's internal structure. We give a short mathematical description of the TOM in Appendix~\ref{sec:TOM} in which we derive the expression for the mantle's rotation vector with respect to the steadily-rotating frame.

At present, the main interest of the results presented here is purely conceptual as our model cannot, at the moment, account for the elasticity of the mantle which induces corrections to the FCN and CW frequencies that are first order in the core ellipticity and therefore dominates over the second order corrections computed here. As it stands now, our model can nevertheless provide a useful baseline for comparison with studies performed in triaxial geometries.

Other possible extensions to the present work include taking into consideration the effects of density stratification inside the core. \citet{Toomre1974} argued that a radial density profile could lead to couple the rotation to inertial modes other than the SO. For the Earth, the inclusion of a magnetic field can also potentially alter the spectrum of inertial modes. This is especially important for the Earth, as the ohmic power dissipation is estimated to be one of the main sources of damping of the FCN \citep{Koot2010}. Finally, the presence of a solid inner core should be taken into account in order to model the \emph{Free Inner Core Nutation} of the Earth. This proves to be a significant numerical challenge as this no longer permits the usage a single set of oblate spheroidal coordinates covering the whole volume of the core. 
Concerning the planet Mars, it is considered (but not yet demonstrated) that there is no inner core and that the main dissipation mechanism at the core-mantle boundary will be induced by the viscosity, also thought to be a very important factor for the Earth [Triana et al. 2020, in prep].
Furthermore, even in the simplified spherical shell geometry, the elliptical nature of the Poincar\'e equation is known to cause singularities in the fluid velocity (and pressure) field \citep{Rieutord2000}. These singularities can be regularised by reintroducing viscosity into the picture, with the well-known numerical limitations that it entails. 

\section*{acknowledgement}
The authors would like to thank T. Van Hoolst for the useful discussions in the writing of this work and the two reviewers, M. Efroimsky and M. Dumberry whose comments helped to improve its presentation significantly. The research leading to the results presented here has received funding from the European Research Council (ERC) under the European Union's Horizon 2020 research and innovation programme (Advanced Grant agreement No. 670874).

\bibliographystyle{abbrvnat}
\bibliography{bibliography}
\newpage

\begin{appendix}

\section{Derivation of the Poincar\'e equation and solutions for the fluid spheroid}
\label{sec:inertialmodesspheroid}
From the momentum Eq.~(\ref{eq:momentumomega}):
\begin{equation}
\sigma\vect{u}+2\hat{\vect{z}}\times\vect{u}+\underbrace{(\grad{p}+\sigma\vect{m}\times\vect{r})}_{\vw}=\vect{0}~,
\label{eq:momentum}
\end{equation}
where we have set $\sigma\equiv i\omega$ for simplicity. Taking the cross product of Eq.~(\ref{eq:momentum}) with $\vz$ and injecting the result back into Eq.~(\ref{eq:momentum}), we find:
\begin{align}
\vect{u}-\frac{4}{\sigma^2}\left(\vz\times(\vz\times\vect{u})\right)-2~\frac{\vz\times\vw}{\sigma^2}+\frac{\vw}{\sigma}&=\vect{0}~,\\
\leftrightarrow~\vect{u}-\frac{4}{\sigma^2}\left(\vz(\vz\cdot\vect{u})-\vect{u}\right)-2~\frac{\vz\times\vw}{\sigma^2}+\frac{\vw}{\sigma}&=\vect{0}~.\label{eq:uu}
\end{align}
From Eq.~(\ref{eq:momentum}), one also has:
\begin{equation}
\vz\cdot\vect{u}=-\frac{(\vz\cdot\vw)}{\sigma}~,
\end{equation}
which can then be used to rewrite the second term of Eq.~(\ref{eq:uu}). This allows one to isolate $\vect{u}$:
\begin{equation}
\vect{u}=\frac{\sigma}{\sigma^2+4}\left(-\vw+\frac{2}{\sigma}(\vz\times\vw)-\frac{4}{\sigma^2}\vz(\vz\cdot\vw)\right)~.
\end{equation}
Eqs.~(\ref{eq:poincareeq}) and (\ref{eq:poincareeqbc}) follow immediately from the condition of incompressibility and the no-penetration condition respectively.

In Sec.~\ref{sec:results}, we compare the inertial modes of the freely rotating two-layer planet with an axi-symmetric fluid core to those of a fluid core in steady rotation around the polar axis, $\vect{\hat{z}}$. Those are the results of Eqs~(\ref{eq:poincareeq}) and (\ref{eq:poincareeqbc}) with $\vect{m}=\vect{0}$. In such case, \citet{Bryan1889} showed that the inertial modes can be written as a sum of product of two \emph{associated Legendre polynomials} (see Eq.~(A.13) of \citep{Rekier2018}). The frequencies of the inertial modes, $\omega$, then derive from the solutions of~:
\begin{equation}
{\text{P}_{\ell}^m}'(x)-\frac{2m x}{\omega(1-x^2)}{\text{P}_{\ell}^m}(x)=0~,
\end{equation}
where $\text{P}_\ell^m(x)=\frac{(-1)^m}{2^\ell\ell!}(1-x^2)^\frac{m}{2}\frac{d^{\ell+m}}{dx^{\ell+m}}(x^2-1)^\ell~,$ denotes the associated Legendre polynomial of degree $\ell$ and order $m$ and where we wrote~:
\begin{equation}
x=\omega\sqrt{\frac{1-e^2}{4-e^2\omega^2}}~,
\end{equation}
with $e\equiv\sqrt{1-\frac{c^2}{a^2}}$ denoting the (geometrical) eccentricity of the fluid core.


\section{The Poincar\'e flow and its associated angular momentum}
\label{sec:poincaréflow}

Eq.~(\ref{eq:upoincare}), gives the general expression of a flow with a uniform vorticity. The scalar function, $\psi$, must be chosen so as to accommodate the no-penetration boundary condition at the CMB, Eq.~(\ref{eq:u.n=0}). Following \citet{Poincare1885}, we do so by operating the following coordinates transform~:
\begin{equation}
\{x,y,z\}\rightarrow\{ax,by,cz\}~,
\end{equation}
where $a$, $b$ and $c$ denote the dimensions of the core (Fig.~\ref{fig:model}). If one defines the rescaled position vector in terms of its cartesian coordinates as $\vect{r}'=(\frac{x}{a},\frac{y}{b},\frac{z}{c})^\text{T}$, the implicit equation of the ellipsoid surface reduces to $|\vect{r}'|^2=1$. The most general flow of uniform vorticity satisfying Eq.~(\ref{eq:u.n=0}) is then~: 
\begin{equation}
\vect{u}=\vect{w}'\times\vect{r}'~,
\end{equation}
where $\vect{w}'$ is a constant vector in space related to the flow vorticity, $\vect{w}$, through the condition that $(\vect{\nabla}\times\vect{u})=2\vect{w}$. In components, this yields
\begin{align}
u^x&=w'^y\frac{a}{c}z-w'^z\frac{a}{b}y=2a^2(\frac{w^yz}{a^2+c^2}-\frac{w^zy}{a^2+b^2})~,\label{eq:px}\\
u^y&=w'^z\frac{b}{a}x-w'^x\frac{b}{c}z=2b^2(\frac{w^zx}{a^2+b^2}-\frac{w^xz}{b^2+c^2})~,\label{eq:py}\\
u^z&=w'^x\frac{c}{b}y-w'^y\frac{c}{a}x=2c^2(\frac{w^xy}{b^2+c^2}-\frac{w^yx}{a^2+c^2})~,\label{eq:pz}
\end{align}
from which we obtain the expression of $\psi$ by comparison with Eq.~(\ref{eq:upoincare})~:
\begin{equation}
\psi=\omega^x~\frac{c^2-b^2}{c^2+b^2}~yz+\omega^y~\frac{a^2-c^2}{a^2+c^2}~zx+\omega^z~\frac{b^2-a^2}{b^2+a^2}~xy~.
\label{eq:psi}
\end{equation}
Using Eqs.~(\ref{eq:px}) to (\ref{eq:psi}), we can finally compute the angular momentum of the core from its definition as measured from the inertial frame~:
\begin{align}
\vect{L}_\text{core}&=\Icore\cdot\Om+\int_\text{core}\rho(\vect{r}\times\vect{u})\\
&=\Icore\cdot(\Om+\vect{w})+\int_\text{core}\rho(\vect{r}\times\grad{\psi})~,
\label{eq:Lcorepoincare}
\end{align}
where $\Icore$ denotes the tensor of inertia of the fluid core and the integrals run over the whole volume of the core. In the cartesian components, this yields~:
\begin{align}
L_\text{core}^x&=\text{A}_f~\Omega_\text{m}^x+\frac{(\text{A}_f+\text{B}_f-\text{C}_f)(\text{C}_f+\text{A}_f-\text{B}_f)}{\text{A}_f}w^x~,\label{eq:Lcx}\\
L_\text{core}^y&=\text{B}_f~\Omega_\text{m}^y+\frac{(\text{B}_f+\text{C}_f-\text{A}_f)(\text{A}_f+\text{B}_f-\text{C}_f)}{\text{B}_f}w^y~,\label{eq:Lcy}\\
L_\text{core}^z&=\text{C}_f~\Omega_\text{m}^z+\frac{(\text{C}_f+\text{A}_f-\text{B}_f)(\text{B}_f+\text{C}_f-\text{A}_f)}{\text{C}_f}w^z~.\label{eq:Lcz}
\end{align}
Using the definitions of the dynamical flattening parameters Eqs.~(\ref{eq:alphafbetaf}), these read~:
\begin{align}
L_\text{core}^x&=\text{A}_f~\Omega_\text{m}^x+\text{A}_f \frac{\left(1-\alpha _f\right) \left(1+\alpha _f-2 \beta_f\right)}{\left(1-\beta _f\right){}^2} w^x\approx\text{A}_f~(\Omega_\text{m}^x+w^x)+\mathcal{O}(\epsilon^2)~,\\
L_\text{core}^y&=\text{B}_f~\Omega_\text{m}^y+\text{B}_f \frac{\left(1-\alpha _f\right) \left(1+\alpha _f+2 \beta _f\right)}{(1+\beta_f)^2} w^y\approx\text{B}_f~(\Omega_\text{m}^y+w^y)+\mathcal{O}(\epsilon^2)~,\\
L_\text{core}^z&=\text{C}_f~\Omega_\text{m}^z+\text{C}_f \frac{\left(1+\alpha _f-2 \beta _f\right) \left(1+\alpha _f+2 \beta_f\right)}{\left(1+\alpha _f\right)^2}w^z\approx\text{C}_f~(\Omega_\text{m}^z+w^z)+\mathcal{O}(\epsilon^2)~.
\end{align} 
where, in the rightmost equalities, we have assumed that $\alpha_f$ and $\beta_f$ are small quantities proportional to a single parameter $\epsilon$. 

\section{Tensor of inertia}
\label{sec:inertiatensor}

The principle moments of inertia of the fluid ellipsoid of equation $\frac{x^2}{a^2}+\frac{y^2}{b^2}+\frac{z^2}{c^2}=1$ and with a homogeneous density $\rho_f$ are~:
\begin{align}
\text{A}_f&=\frac{4\pi}{3}\rho_f~abc~\frac{(b^2+c^2)}{5}~,\label{eq:Af}\\
\text{B}_f&=\frac{4\pi}{3}\rho_f~abc~\frac{(a^2+c^2)}{5}~,\label{eq:Bf}\\
\text{C}_f&=\frac{4\pi}{3}\rho_f~abc~\frac{(a^2+b^2)}{5}~.\label{eq:Cf}
\end{align}
After introduction into the definitions of the dynamical flattening parameters Eqs.~(\ref{eq:alphafbetaf}), we find~:
\begin{align}
\alpha_f&=\frac{a^2+b^2-2c^2}{a^2+b^2+2c^2}~,\label{eq:alphafabc}\\
\beta_f&=\frac{a^2-b^2}{a^2+b^2+2c^2}~.\label{eq:betafabc}
\end{align}
Substituting Eqs.~(\ref{eq:alphafabc}) and (\ref{eq:betafabc}) into Eqs.~(\ref{eq:Af}) to (\ref{eq:Cf}), yields~:
\begin{align}
\text{A}_f&=\frac{8\pi}{15} a^5\rho_f\frac{\sqrt{(1-\alpha_f ) (\alpha_f -2 \beta_f +1)}}{(\alpha_f +2 \beta_f +1)^2} (1-\beta_f )~,\\
\text{B}_f&=\frac{8\pi}{15} a^5\rho_f\frac{\sqrt{(1-\alpha_f ) (\alpha_f -2 \beta_f +1)}}{(\alpha_f +2 \beta_f +1)^2} (1+\beta_f )~,\\
\text{C}_f&=\frac{8\pi}{15} a^5\rho_f\frac{\sqrt{(1-\alpha_f ) (\alpha_f -2 \beta_f +1)}}{(\alpha_f +2 \beta_f +1)^2} (1+\alpha_f )~.
\end{align}
We can obtain the expressions of the moments of inertia of the mantle by working directly from the above expressions. If we define the ratio between the semi-major axes of the core and of the whole planet as $\eta$, the principle moments of inertia of the mantle read~:
\begin{align}
\text{A}_m&=\frac{8\pi}{15} a^5\rho_m\left(\frac{\sqrt{(1-\alpha) (\alpha-2 \beta+1)}}{(\alpha+2 \beta+1)^2} \frac{(1-\beta)}{\eta^5}-\frac{\sqrt{(1-\alpha_f ) (\alpha_f -2 \beta_f +1)}}{(\alpha_f +2 \beta_f +1)^2} (1-\beta_f )\right)~,\\
\text{B}_m&=\frac{8\pi}{15} a^5\rho_m\left(\frac{\sqrt{(1-\alpha) (\alpha-2 \beta+1)}}{(\alpha+2 \beta+1)^2} \frac{(1+\beta)}{\eta^5}-\frac{\sqrt{(1-\alpha_f ) (\alpha_f -2 \beta_f +1)}}{(\alpha_f +2 \beta_f +1)^2} (1+\beta_f )\right)~,\\
\text{C}_m&=\frac{8\pi}{15} a^5\rho_m\left(\frac{\sqrt{(1-\alpha) (\alpha-2 \beta+1)}}{(\alpha+2 \beta+1)^2} \frac{(1+\alpha)}{\eta^5}-\frac{\sqrt{(1-\alpha_f ) (\alpha_f -2 \beta_f +1)}}{(\alpha_f +2 \beta_f +1)^2} (1+\alpha_f )\right)~.
\end{align}

\section{The steadily rotating frame and the Tilt-Over mode}
\label{sec:TOM}

On Fig.~\ref{fig:Ekinr}, we have represented the ratio between the kinetic energy density of the mantle and that of the fluid core as measured in the \emph{steadily rotating frame}, which corresponds to the frame rotating with a constant velocity around the mean axis of rotation of the planet. In their Appendix C, \citet{Triana2019} gave some details on how the values of the quantities computed in that frame relate to their value in the mantle frame. We now provide a more detailed explanation.

We can think of the $z$-axis in the inertial frame, which we denote by $\vect{\hat{z}}_\text{IF}$ as the mean rotation axis of the planet perpendicular to the equator (at J2000). The angular velocity of the mantle relative to the steadily rotating frame therefore reads (assuming $\Omega_0=1$):
\begin{equation}
\vect{\Omega}_{\text{m}|\text{SRF}}=(\vect{\hat{z}}+\vect{m})-\vect{\hat{z}}_\text{IF}~,
\end{equation}
where $\vect{\hat{z}}$ denotes the instantaneous $z-$axis of the mantle frame. In order to write this in coordinates, we need to express $\vect{\hat{z}}_\text{IF}$ in the mantle's coordinates. We can do this by using the three Euler angles that relate the mantle to the steadily rotating frame. We use the following convention:
\begin{equation}
\underbrace{\left(
\begin{array}{c}
\cdot\\\cdot\\\cdot\\
\end{array}
\right)}_\text{mantle coordinates}=\underbrace{R_z(\gamma)\cdot R_y(\beta)\cdot R_x(\alpha)\cdot R_z(t)}_T
\underbrace{\left(
\begin{array}{c}
\cdot\\\cdot\\\cdot\\
\end{array}
\right)}_\text{inertial frame coordinates}~,
\end{equation}
where the $R_i$ denotes the ordinary (three dimensional) rotation matrices around the axis $i$. The arguments of these matrices define the Euler angles $\alpha$, $\beta$ and $\gamma$ which are all harmonic functions of $t$ with frequency $\omega$, and should not be confused with the flattening parameters of Eqs.~(\ref{eq:alphabeta}). From the transformation matrix, $T$, we can construct the antisymmetric rotation tensor:
\begin{equation}
\vect{R}\equiv\frac{dT}{dt}\cdot T^\text{T}~.
\end{equation}
The dual vector of $\vect{R}$ then defines the angular velocity of the rotation (see \emph{e.g.}, p.~46 of \citet{lai2010}) which must be equal to the angular velocity of the mantle. In components and assuming that $\alpha$, $\beta$ and $\gamma$ are small, this gives:
\begin{equation}
\begin{cases}
m^x=i\omega\alpha-\beta\\
m^y=i\omega\beta-\alpha\\
m^z=i\omega\gamma~.
\end{cases}
\label{eq:eulerm}
\end{equation}
In order to obtain the components of $\vect{\hat{z}}_\text{IF}$ in the mantle frame, we use the transformation matrix:
\begin{equation}
T\cdot 
\left(
\begin{array}{c}
0\\ 0 \\ 1\\
\end{array}
\right)=\left(
\begin{array}{c}
-\beta\\ \alpha \\ 1\\
\end{array}
\right)~.
\end{equation}
Finally, after solving Eq.~(\ref{eq:eulerm}) for $\alpha$ and $\beta$ we obtain the expression of the angular velocity of the mantle relative to the steadily rotating frame in the mantle's coordinates (in terms of $m^+$ and $m^-$):
\begin{equation}
\vect{\Omega}_{\text{m}|\text{SRF}}=\left(
\begin{array}{c}
\frac{m^+\omega}{\sqrt{2}(1-\omega)}+\frac{m^-\omega}{\sqrt{2}(1+\omega)}\\ 
\frac{im^+\omega}{\sqrt{2}(1-\omega)}-\frac{im^-\omega}{\sqrt{2}(1+\omega)} \\ 
m^z\\
\end{array}
\right)+\emph{c.c.}
\label{eq:OmegamSRF}
\end{equation}
Eq.~(\ref{eq:OmegamSRF}) presents a clear resonance when $\omega=\pm1$. These values can be interpreted as new eigenfrequencies which are added to the spectrum when we solve the first two equations of Eq.~(\ref{eq:eulerm}) for $\alpha$ and $\beta$ simultaneously with Eqs.~(\ref{eq:Mmxmy}). They correspond to the motion known as the \emph{Tilt-Over mode} which has a purely diurnal frequency independent to the interior structure of the planet.

\end{appendix}

\end{document}